\documentclass[%
 aps,
 prb,
 amsmath,amssymb,
reprint,%
]{revtex4-2}

\usepackage{graphicx}
\usepackage{dcolumn}
\usepackage{bm}

\usepackage[utf8]{inputenc}
\usepackage[T1]{fontenc}
\usepackage{mathptmx}
\usepackage{multirow}
\usepackage{siunitx}
\usepackage{float}
\usepackage{color}
\usepackage{comment}

\begin{document}

\preprint{APS}

\title{Controlling Domain-Wall Nucleation in Ta/CoFeB/MgO Nanomagnets via Local Ga\textsuperscript{+} Ion Irradiation}
\author{Simon Mendisch}%
\email{simon.mendisch@tum.de}
\affiliation{Department of Electrical and Computer Engineering, Technical University of Munich, Arcisstr. 21, 80333, Munich, Germany
}%
\author{Fabrizio Riente}
 \email{fabrizio.riente@polito.it}
 \affiliation{Department of Electronics and Telecommunications Engineering, Politecnico di Torino, 10129, Turin, Italy.}
\author{Valentin Ahrens}%
\affiliation{Department of Electrical and Computer Engineering, Technical University of Munich, Arcisstr. 21, 80333, Munich, Germany
}%
\author{Luca Gnoli}
\affiliation{Department of Electronics and Telecommunications Engineering, Politecnico di Torino, 10129, Turin, Italy.}%

\author{Michael Haider}%
\affiliation{Department of Electrical and Computer Engineering, Technical University of Munich, Arcisstr. 21, 80333, Munich, Germany
}%
\author{Matthias Opel}%
\affiliation{Bavarian Academy of Sciences, Walther-Meißner-Straße 8, 85748,  Garching, Germany
}%
\author{Martina Kiechle}%
\affiliation{Department of Electrical and Computer Engineering, Technical University of Munich, Arcisstr. 21, 80333, Munich, Germany
}%
\author{Massimo Ruo Roch}
\affiliation{Department of Electronics and Telecommunications Engineering, Politecnico di Torino, 10129, Turin, Italy.}%
\author{Markus Becherer}%
\affiliation{Department of Electrical and Computer Engineering, Technical University of Munich, Arcisstr. 21, 80333, Munich, Germany
}

\date{\today}

\begin{abstract}
Comprehensive control of the domain wall nucleation process is crucial for spin-based emerging technologies ranging from random-access and storage-class memories over domain-wall logic concepts to nanomagnetic logic. In this work, focused Ga\textsuperscript{+} ion-irradiation is investigated as an effective means to control domain-wall nucleation in Ta/CoFeB/MgO nanostructures. We show that analogously to He\textsuperscript{+} irradiation, it is not only possible to reduce the perpendicular magnetic anisotropy but also to increase it significantly, enabling new, bidirectional manipulation schemes. First, the irradiation effects are assessed on film level, sketching an overview of the dose-dependent changes in the magnetic energy landscape. Subsequent time-domain nucleation characteristics of irradiated nanostructures reveal substantial increases in the anisotropy fields but surprisingly small effects on the measured energy barriers, indicating shrinking nucleation volumes. Spatial control of the domain wall nucleation point is achieved by employing focused irradiation of pre-irradiated magnets, with the diameter of the introduced circular defect controlling the coercivity. Special attention is given to the nucleation mechanisms, changing from a Stoner-Wohlfarth particle's coherent rotation to depinning from an anisotropy gradient. Dynamic micromagnetic simulations and related measurements are used in addition to model and analyze this depinning-dominated magnetization reversal.

\end{abstract}

\maketitle

 \section{Introduction}
Magnetic nanostructures based on Cobalt-Iron-Boron/Magnesium-oxide (CoFeB/MgO) thin films, with and without perpendicular magnetic anisotropy (PMA), play a vital role in many emerging technologies, from magnetic tunnel-junction based sensors over non-volatile storage technologies, towards domain-wall and nanomagnetic logic applications\cite{garello2018sot,riente2020ta,garello2019manufacturable,sakhare2018enablement,xie2015hysteresis}. Especially logic applications necessitate precise control of the magnetic energy landscape to nucleate, propagate and pin/depin domain-walls — a level of control that remains a significant challenge \cite{manfrini2018interconnected, mendisch2020perpendicular}.
As widely established semiconductor technology with unmatched spatial resolution and a wide tuning range, ion irradiation is ideally suited to address these issues \cite{fassbender2008magnetic}. It offers a realistic perspective to modify magnetic properties with nanometer precision. So far, studies on the irradiation effects on CoFeB/MgO have mainly been restricted to film level investigations and light (He\textsuperscript{+}) ions \cite{devolder2013irradiation, herrera2015controlling, diez2019enhancement}. In this work, we investigate the usage of heavier Ga\textsuperscript{+} ions in an attempt to create artificial nucleation centers (ANC) in Ta/CoFeB/MgO nanomagnets with PMA, employing localized ion irradiation (not implantation), thus controlling domain wall (DW) nucleation. Gallium ions are chosen, as they are known to reduce the anisotropy in crystalline multilayer systems effectively \cite{franken2011domain}. Heavier atoms, furthermore, can be stopped much more effectively, reducing potential damage to underlying layers. The dose-dependent irradiation effects are first evaluated on film level, probing material parameter and domain configurations, before the focus is shifted towards the irradiation of nanostructures and time-domain measurements. We thereby explain the different time-dependent DW nucleation probabilities, from which information regarding nucleation mechanisms (coherent rotation or depinning) and irradiation effects are derived. Unitizing this analysis, we employ irradiation at the magnet's centers to control the switching fields and force DW nucleation via depinning instead of coherent rotation.

\section{Fabrication and Characterization}
\subsection{Device Fabrication} \label{sec:Fab}
The magnetic thin film analyzed in this work is based on a Ta/CoFeB/MgO/Ta sandwich structure with a Co\textsubscript{20}Fe\textsubscript{60}B\textsubscript{20} alloy target and nominal thicknesses of Ta\textsubscript{2}/CoFeB\textsubscript{1}/MgO\textsubscript{2}/Ta\textsubscript{3} (numbers given in \SI{}{\nano\meter}). The film is deposited at room temperature via confocal RF-magnetron sputtering (base pressure < \SI{2e-7}{\milli\bar}) onto silicon $\left\langle 100 \right\rangle$  substrates, topped by a thermal oxide (thickness $\approx$ \SI{50}{\nano\meter}). The individual materials are deposited at a constant working pressure of \SI{4}{\micro\bar} ($\approx$ \SI{3}{\milli Torr}) with MgO as an exception (< \SI{1}{\micro\bar}). The target power density is \SI{0.5}{\watt\per\square\centi\meter} for all materials. Post deposition annealing ( \SI{250}{\celsius}, N\textsubscript{2} atmosphere) is used to set the effective anisotropy to the desired value of $\approx$ \SI{1.3E5}{\joule\per\cubic\meter}.
The stack is subsequently structured via focused ion beam (FIB) lithography (using PMMA as a positive ion-beam resist) to realize the designed test-structures. The lithography profile is inverted, after PMMA development, by depositing a \SI{5}{\nano\meter} thick Ti hard-mask and removing the residual PMMA in a lift-off process. Finally, the non-masked areas are physically etched via Ar\textsuperscript{+} ion-beam etching ($E= \SI{350}{\electronvolt}$). To generate ultra-short magnetic field pulses, on-chip field coils are placed around the structures via conventional optical contact lithography together with the deposition of a Cu metal layer ($\approx$ \SI{750}{\nano\meter}) and a second lift-off process. 
The Ga\textsuperscript{+} ion-irradiation, changing the magnetic properties, is carried out using a $\SI{50}{\kilo\electronvolt}$ focused-ion-beam (FIB) microscope (Micrion 9500ex) with a spatial resolution (beam diameter) of $\approx \SI{10}{\nano\meter}$. For large areas, the beam is de-focused to achieve homogeneous irradiation results.

\subsection{Magneto-Optical Imaging} \label{sec:Imaging}

\begin{figure}[b]
\centering
\includegraphics[width=0.5\textwidth]{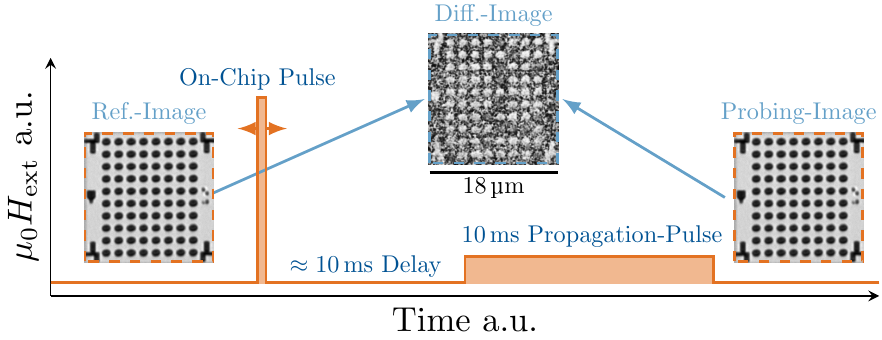}
\caption{Sketch of the employed measurement scheme, illustrating the procedures in a time-line. First, a reference image of the saturated magnet array is taken. Subsequently, the on-chip pulse with varying widths of $t_{\mathrm{p}}=$\SI{10}{\nano\second} to \SI{100}{\micro\second} is triggered before a second, millisecond long pulse is used to propagate remaining DWs and complete the reversal process. After the propagation pulse, a second image is taken, which is subtracted from the reference image. The final difference image is then used for the later analysis. This procedure is repeated multiple times to gain statistical data for the magnets in the image frame. }
\label{fig:Measuremnts}  
\end{figure}
The magnetic nanostructures are characterized via Wide-field Kerr-microscopy (WMOKE), both for the quasi-static case as well as for time-domain measurements.
Static coercivities are obtained by  merely applying a stair-case field profile with images taken at every step. In a later data-processing step, the coercivities of the individual magnets can be derived from the respective brightness changes in the images. Reliable time-domain measurements, however, require a more sophisticated measurement scheme. Ultra short magnetic field pulses are generated via (single winding) on-chip coils, which are bonded to pulse discharge-capacitors on the high-side, and a low-side switch. This switch is driven by a fast gate-driver and is addressed with an Agilent 81111A Pulse Generator. To measure the nucleation probability $p_{\mathrm{nuc}}$ of the individual magnets at short timescales, the on-chip field pulses in the \SI{}{\nano\second}-range are synchronized with the image acquisition of a high dynamic-range sCMOS camera. Figure~\ref{fig:Measuremnts} depicts a rough sketch of this imaging procedure. After the initial saturation of the magnets, a first reference image is taken. Consecutively, the on-chip pulse is triggered to nucleate a DW.
A second propagation pulse with a low amplitude ($\approx \SI{3}{\milli\tesla} $) is generated thereafter via an external electromagnet to ensure the complete magnetization reversal (and thus optical detection) upon the nucleation of a DW. The second image, taken after the pulses, enables differential imaging in a later data-processing step. This procedure is repeated multiple times to retrieve the nucleation probability at a given field amplitude and pulse width. 
Examples of the probability evolution are displayed in Fig.~\ref{fig:NukePos} (a).
The advantage of this procedure, compared to laser-based approaches, is the ability to probe large numbers of magnets simultaneously by facilitating image recognition to detect and label all magnets within the image frame.

\begin{figure*}[t]
\centering
\includegraphics[width=1\textwidth]{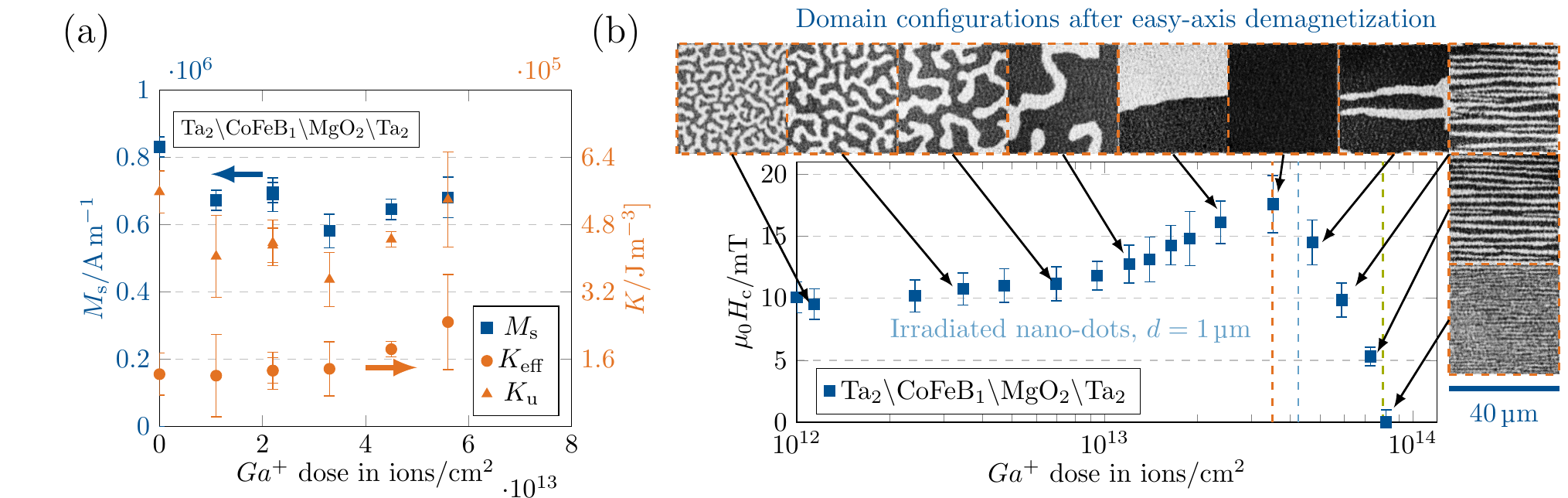}
\caption{The plot in (a) depicts the measured material parameter  $M_{\mathrm{s}}$ and $K_{\mathrm{eff}}$ of the film, together with the calculated uniaxial anisotropy term $K_{\mathrm{u}} = K_{\mathrm{eff}}- \frac{1}{2}\mu_0 M_{\mathrm{s}}^2$ versus the applied ion dose. In (b), the Ga\textsuperscript{+} dose-dependent coercivity evolution of circular Ta\textsubscript{2}/CoFeB\textsubscript{1}/MgO\textsubscript{2} nano-magnets ($d = \SI{1}{\micro\meter}$) is depicted. The error bars indicate the raw FWHM switching field distribution of 80 magnets each. The surrounding domain images display dose correlated domain-patterns imaged on the same film after irradiation and easy-axis demagnetization. The colored dashed lines serve as markers, indicating doses used in more dialed analysis.  }
\label{fig:HvsKeff}  
\end{figure*}
\section{Results \& Discussion}
\subsection{Areal Irradiation and Static Measurements}
To understand and interpret the irradiation dependent changes in the domain wall dynamics of nanostructures, we first analyze the irradiation effects on film level. This allows probing the essential material parameter ($M_{\mathrm{s}}$ and $K_{\mathrm{eff}}$) via comparatively simple though error-prone magnetometer measurements. The material parameters are extracted from SQUID and VSM-magnetometer loops. $K_{\mathrm{eff}}$ is thereby approximated from the hard-axis loops via the area method \cite{johnson1996magnetic}. The uniaxial anisotropy constant $K_{\mathrm{u}}$, necessary for the micromagnetic simulations, is calculated as $K_{\mathrm{u}} = K_{\mathrm{eff}}+ \frac{1}{2}\mu_0 M_{\mathrm{s}}^2$. Figure~\ref{fig:HvsKeff}  (a) depicts the  irradiation-induced changes in $M_{\mathrm{s}}$ as well as $K_{\mathrm{eff}}$ with increasing ion dose. Similar to reports on the He\textsuperscript{+} irradiation of Ta/CoFeB/MgO films, a decrease in saturation magnetization accompanied by an increase in anisotropy is observed. Figure~\ref{fig:HvsKeff} (b) furthermore depicts the irradiation dependent static coercivities ($H_{\mathrm{c}}$) of circular nano-dots ($d= \SI{1}{\micro\meter}$) and respective (dose matched) domain images. The changes in the coercivities and domain sizes enable a more detailed though qualitative assessment of the shifts in the anisotropy landscape as both scale $\propto K_{\mathrm{eff}}$. The data points display the center of the respective switching field distribution (SFD), with the error bars indicating the full width at half maximum (FWHM). Thereby, 80 magnets are probed for each ion dose. The coercivity and thus $K_{\mathrm{eff}}$ initially increases for low and medium doses and only starts to fall off at doses higher $\approx \SI{3.5E13}{}\, \mathrm{ions/cm^2}$ with the domain size dropping below the resolution limit at a dose of $\approx \SI{8E13}{}\, \mathrm{ions/cm^2}$. The magnets cross the single-domain threshold at this point.  It has to be noted that the apparent decrease in $K_{\mathrm{eff}}$ above $\approx \SI{3.5E13}{}\, \mathrm{ions/cm^2}$ could not be replicated via corresponding magnetometer measurements. This fact might, however, be explained by the macroscopic nature of the magnetometer measurements, complicating the detection of small changes in the anisotropy landscape. Explaining the non-monotonic evolution of $K_{\mathrm{eff}}$ is difficult without a detailed stoichiometric analysis, and therefore no comprehensive explanation can be given. However, as with He\textsuperscript{+} irradiation \cite{nembach2020tuning,devolder2013irradiation}, the behavior might be explained by the respective atomic weights of the different elements inside the stack, giving the Ga\textsuperscript{+} ions a much larger probability to interact with the heavy Ta rather than with the comparatively light Fe, Co, or O atoms. Since Tantalum is known for its large magnetic dead layer in contact with ferromagnets, we assume intermixing at the Ta/CoFeB interface to be the dominant cause for the decrease in $M_{\mathrm{s}}$\cite{nembach2020tuning,devolder2013irradiation}. A possible explanation for the non-monotonicity in $K_{\mathrm{eff}}$ could be that due to this reduced interaction probability, the damage to the CoFeB/MgO interface and thus $K_{\mathrm{u}}$ only becomes relevant at much higher doses\cite{nembach2020tuning,devolder2013irradiation}. Closely related to this is the likely accumulation of Tantalum atoms at the CoFeB/MgO interface, also strongly affecting the anisotropy \cite{miyakawa2013impact}. An interesting observation related to the anisotropy decrease is the formation of highly ordered stripe domains at high ion doses, indicating changes in more than the primary material parameter. 
This is in line with reports on the increase of the interfacial Dzyaloshinskii-Moriya interaction upon the irradiation of Ta/CoFeB/Mg films  \cite{diez2019enhancement}.

\subsection{Controlling the Magnetization Reversal}
\begin{figure}[h]
\centering
\includegraphics[width=0.5\textwidth]{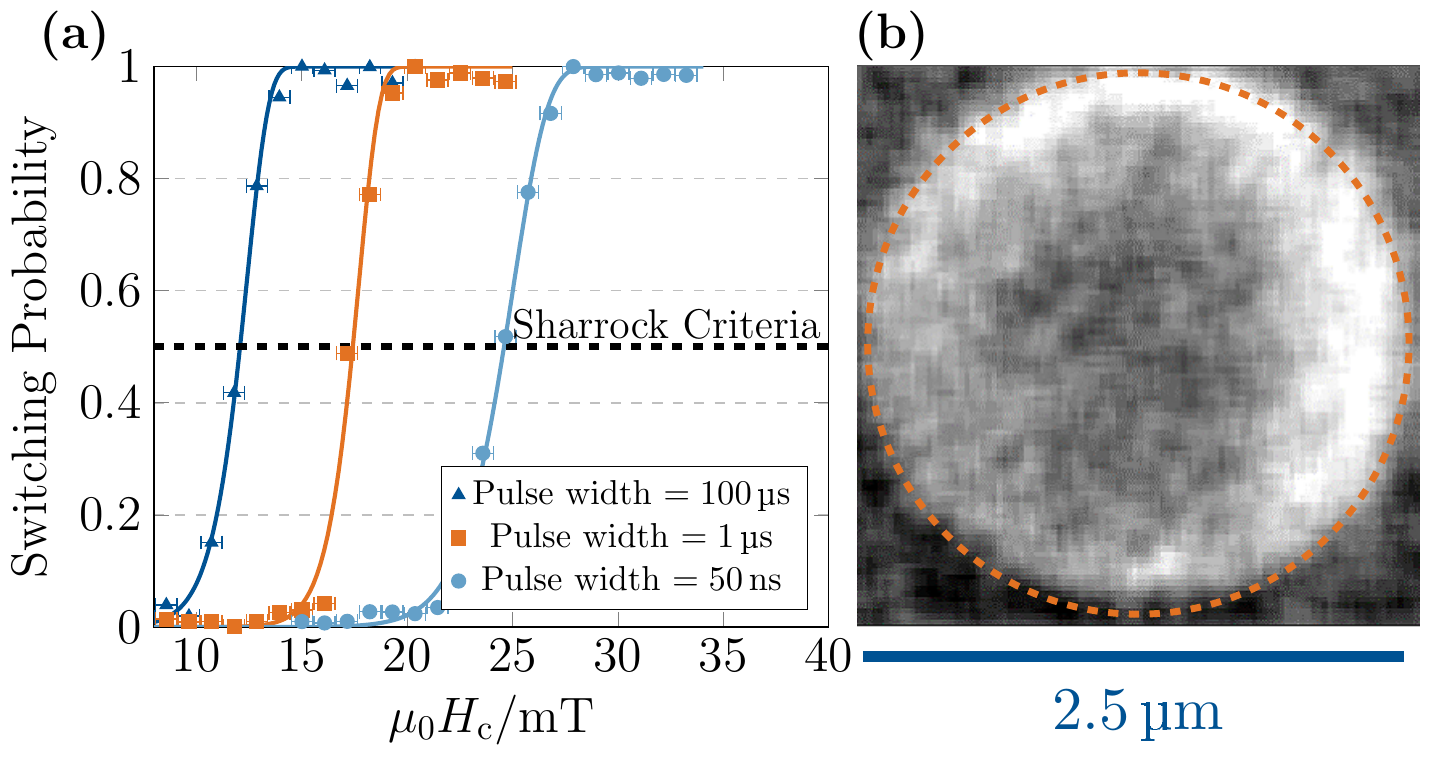}
\caption{Plot (a) depicts the measured, field-dependent, nucleation probabilities of a \SI{1}{\micro\meter} nanomagnet for different pulse widths.
The image in (b) depicts 1000 superimposed differential WMOKE images of the nucleation events inside a \SI{2.5}{\micro\meter} wide nano-dot. The areas of increased brightness indicate higher nucleation probabilities. Nucleation is achieved using \SI{20}{\nano\second} long pulses without consecutive propagation pulses.}
\label{fig:NukePos}  
\end{figure}

We have already shown that the magnets' coercivities can be effectively tailored by adjusting the ion dose. However,
static measurements only provide limited insight into the reversal mechanisms and are not suited to derive relevant conclusions. Therefore, we attempt a characterization of the irradiation dependent reversal process by probing the time-dependent magnetization reversal. For this purpose, we provide a sample-base of at least 40 magnets per data-point, reducing the effects of statistical outliers. Contrary to the distribution of the demagnetizing fields, DWs in CoFeB/MgO nano-magnets usually nucleate at the nanostructures' edges due to an etch-damage induced lowering of $K_{\mathrm{eff}}$ \cite{zhang2018extrinsic, zhang2018domain}. To validate this assumption for the test structures, \SI{20}{\nano\second} long magnetic field pulses are used to nucleate DWs in circular nano-disks with a diameter of \SI{2.5}{\micro\meter} with the goal to locate the nucleation sights via repeated differential WMOKE imaging. Figure~\ref{fig:NukePos} (b) displays the combined differential WMOKE image of ten different disks with a total of 1000 superimposed images to qualitatively show the local nucleation probability. Bright areas thereby indicate an increased DW nucleation probability. 

The image indicates the accumulation of nucleation events at the edges of the disks, while an inhomogeneity in the applied on-chip fields most likely explains the asymmetry towards the right side. Images of single nucleation events and a sanity check without nucleation can be found in the supplementary information (SI). The conformation of nucleation from the edges has severe implications. Instinctively, one would expect the nucleation to occur at points with strong demagnetizing fields, i.e., the center of the magnet. However, the demagnetizing fields are lowest at the edges, leading to the conclusion that the reduction in $K_{\mathrm{eff}}$  must be significantly larger than the anisotropy variations in the magnets' center. Furthermore, the question arises, whether the DW nucleation occurs via coherent rotation according to the \textit{Stoner–Wohlfarth} model or by depinning from an area with easy-plane anisotropy \cite{stoner1948mechanism, choe2006direct}. This can be resolved by considering the time evolution of both processes. The rotation fields scale over time according to the well established \textit{Sharrock} formalism based on an Arrhenius switching model of a \textit{Stoner–Wohlfarth} particle and can be expressed by
\begin{equation}\label{Eq:sharrock}
H_{\mathrm{nuc}} = H_{\mathrm{s0}} \left\lbrace1- \left[ \frac{k_{\mathrm{B}} T}{E_{\mathrm{0}}} \ln \left(  \frac{f_{\mathrm{0}} t_{\mathrm{p}}}{\ln (2)} \right)\right]^{\frac{1}{2}}  \right\rbrace  \, ,
\end{equation}
where $H_{\mathrm{s0}}$ is the switching field at $\SI{0}{\kelvin}$, $f_{\mathrm{0}} $ is the attempt frequency ($\approx \SI{1E9}{\hertz}$), and $E_{\mathrm{0}}$ is the energy barrier without applied field  \cite{sharrock1999measurement, stoner1948mechanism, wernsdorfer1997experimental}.
In contrast, the time necessary for a DW to overcome the anisotropy gradient and depin can be derived from the related Néel–Brown theory and scales according to 
\begin{equation}\label{Eq:Depinn}
\tau = f_{0}^{-1} \exp \left[ \frac{M_{\mathrm{s}}V_{\mathrm{a}}}{k_{\mathrm{B}} T} (H_{\mathrm{d}}-H) \right] \, ,
\end{equation}
with $V_{\mathrm{a}}$ as the activation volume and $H_{\mathrm{d}}$ as the depinning field at $\SI{0}{\kelvin}$ \cite{brown1963thermal,neel1949theorie,choe2006direct}.
By characterizing the switching fields over a wide range of different timescales (pulse widths) and comparing the evolution to the models in Eq.~\eqref{Eq:sharrock} and Eq.~\eqref{Eq:Depinn}, it is possible to gather detailed information about the switching mechanisms.
Figure~\ref{fig:sharrock} displays the pulse-width dependent nucleation fields of the circular nano-dots with a diameter of \SI{1}{\micro\meter}. The measurements cover timescales ranging from the quasi-static case down to \SI{10}{\nano\second}. The data points resemble the center of the distribution, with the error-bars again displaying the FWHM. The nucleation field $H_{\mathrm{nuc}}$ is furthermore defined according to the \textit{Sharrock} formalism, as the field with a switching probability $p_{\mathrm{nuc}}\geq \SI{50}{\percent}$.
Figure~\ref{fig:NukePos} (a) shows a series of exemplary nucleation probability measurements for different pulse widths with the \textit{Sharrock} criteria indicated as a dashed line. The plot furthermore depicts corresponding fits according to the Arrhenius switching model with the probability $p_{\mathrm{nuc}} =1-\exp(\frac{-t_{\mathrm{p}}}{\tau_{\mathrm{nuc}}})$, with $\tau_{\mathrm{nuc}}$ as the inverse of the nucleation rate \cite{wernsdorfer1997experimental}.
\subsubsection{Nucleation by Coherent Rotation}
We first consider the pristine magnets and compare the data to the aforementioned nucleation and depinning dominated models.
The nucleation fields show good agreement with the numerical fits according to the \textit{Sharrock} equation, displayed as black lines; the fitting parameters converge to $H_{\mathrm{s0}}= \SI[separate-uncertainty = true]{36.83(165)}{\milli\tesla} $ and  $\frac{E_{\mathrm{0}}}{k_{\mathrm{B}} T}  = (30.98 \pm 2.9) $. Additionally, we attempt to fit equation Eq.~\eqref{Eq:Depinn} analytically by minimizing its cumulative error-function utilizing a linearized least-squares problem \cite{SI}. However, an acceptable solution (displayed as a dotted line) is only obtained excluding pulse widths $< \SI{1}{\micro\second}$, thus arguing against depinning as the primary DW nucleation mechanism, at least for very short timescales. Interestingly, however, at long time scales ($> \SI{10}{\micro\second}$) depinning from the nucleation sites, could very well be the limiting factor, thus explaining the apparent underestimation of the \textit{Sharrock} fits at quasi-static fields.
\begin{figure}[h]
\centering
\includegraphics[width=0.5\textwidth]{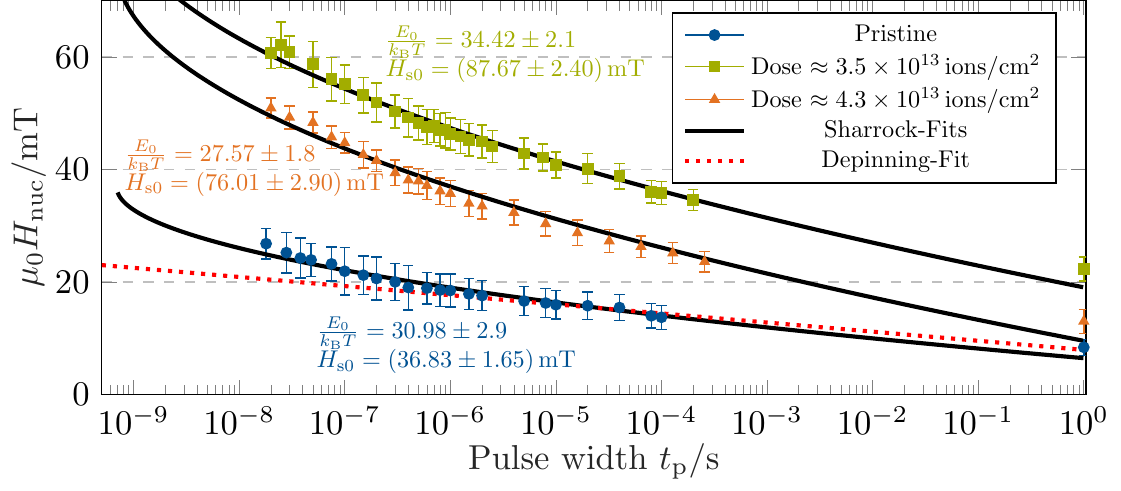}
\caption{Calculated nucleation fields ($H_{\mathrm{nuc}}$) depending on the applied
pulse width. The individual data points display the center of the SFDs with the error bars displaying the FWHM. The corresponding \textit{Sharrock} fits are illustrated as black lines.
A fit, assuming depinning mediated nucleation according to equation~\eqref{Eq:Depinn} for the pristine magnets is illustrated in red.}
\label{fig:sharrock}  
\end{figure}
The question now arises whether Ga\textsuperscript{+} irradiation not only increases the effective anisotropy of the disk's core but whether its effect on the pre-damaged edges is different. Therefore, Fig.~\ref{fig:sharrock} also displays the time evolution of nano-disks homogeneously irradiated with a dose of $\SI{3.5E13}{}\, \mathrm{ions/cm^2}$ and $\SI{4.3E13}{}\, \mathrm{ions/cm^2}$. The doses are chosen to probe the peak of the static coercivity increase as well as a position within the downward slope. For better illustration, the doses are marked, in their respective colors, as dashed lines in Fig.~\ref{fig:HvsKeff} (b). The slopes, again, indicate nucleation by coherent rotation as the dominant mechanism. From the corresponding \textit{Sharrock} fits, we derive the energy barriers to be $\frac{E_{\mathrm{0}}}{k_{\mathrm{B}} T}  = (34.42 \pm 2.1) $ and $\frac{E_{\mathrm{0}}}{k_{\mathrm{B}} T}  = (27.57 \pm 1.8) $, respectively. The fields at which these barriers become zero are determined to be $H_{\mathrm{s0}}= \SI[separate-uncertainty = true]{87.67(240)}{\milli\tesla} $ and $H_{\mathrm{s0}}= \SI[separate-uncertainty = true]{76.01(290)}{\milli\tesla}$. The energy barrier can be roughly modeled as $E_{\mathrm{0}} \approx K_{\mathrm{eff}}V_{\mathrm{nuc}}$ with $V_{\mathrm{nuc}}$ as the nucleation volume (not to be confused with the activation volume $V_{\mathrm{a}}$)\cite{sharrock1999measurement}. The nucleation field at $\SI{0}{\kelvin}$, on the other hand, is equal to the anisotropy field $H_{\mathrm{anis}} \approx \frac{2K_{\mathrm{eff}}}{M_{\mathrm{s}}}$ \cite{sharrock1999measurement, stoner1948mechanism}. Comparing the derived parameters of the pristine magnets with those irradiated, two distinct observations become apparent. The intrinsic switching field $H_{\mathrm{s0}}$ scales in accordance with the increase in anisotropy and the reduction of $M_{\mathrm{s}}$. The energy barrier $\frac{E_{\mathrm{0}}}{k_{\mathrm{B}} T}$, however, increases (if at all) only marginally, despite the increase in $K_{\mathrm{eff}}$. Of course, the conducted measurements are limited in scope and only allow for a cautious interpretation of this unexpected result. Irradiation induced reductions in the nucleation volumes $V_{\mathrm{nuc}}$ could compensate for or even surpass the increase in anisotropy. However, this would conflict with the current, simplistic picture of a homogeneous change in the material parameter of an effective medium. The analyzed doses translate to one Ga\textsuperscript{+} ion roughly every \SI{2}{\nano\meter} if applied homogeneously. The focused ion beam is, thereby, scanned horizontally (line-wise) across the magnets with a constant speed. While the horizontal and vertical ion spacings are assumed to be reasonably constant, they are not expected to be the same, as the horizontal lines must be stitched together vertically. This artificial lattice could account for the reduced nucleation volumes. However, additional studies are needed to give a more detailed answer.
Attempts to determine the local nucleation probability as for the pristine magnets were unsuccessful, as the high nucleation fields result in very high DW velocities, leading to complete reversals already within a fiew \SI{}{\nano\second}.


\subsubsection{Nucleation by Depinning}
\begin{figure}[t]
\centering
\includegraphics[width=0.5\textwidth]{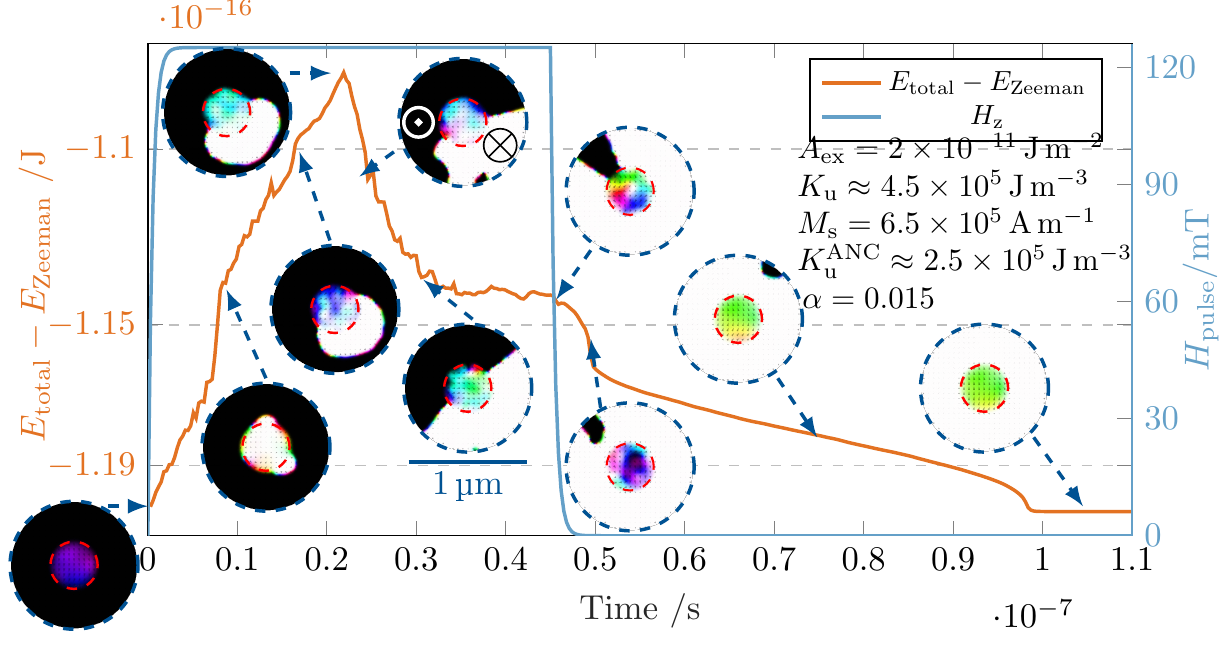}
\caption{Plot of the simulated magnetization reversal process, depicting a \SI{1}{\micro\meter} circular nanomagnet with centered ANC ($ d = \SI{400}{\nano\meter}$). The graph displays the combined magnetic energies (excluding the \textit{Zeeman} term) in combination with snapshots of the domain structure at relevant points. The assumed material parameters are listed in the plot. Further information about the simulations can be found in the supplementary information.}
\label{fig:Laplace}  
\end{figure}

\begin{figure}[b]
\centering
\includegraphics[width=0.5\textwidth]{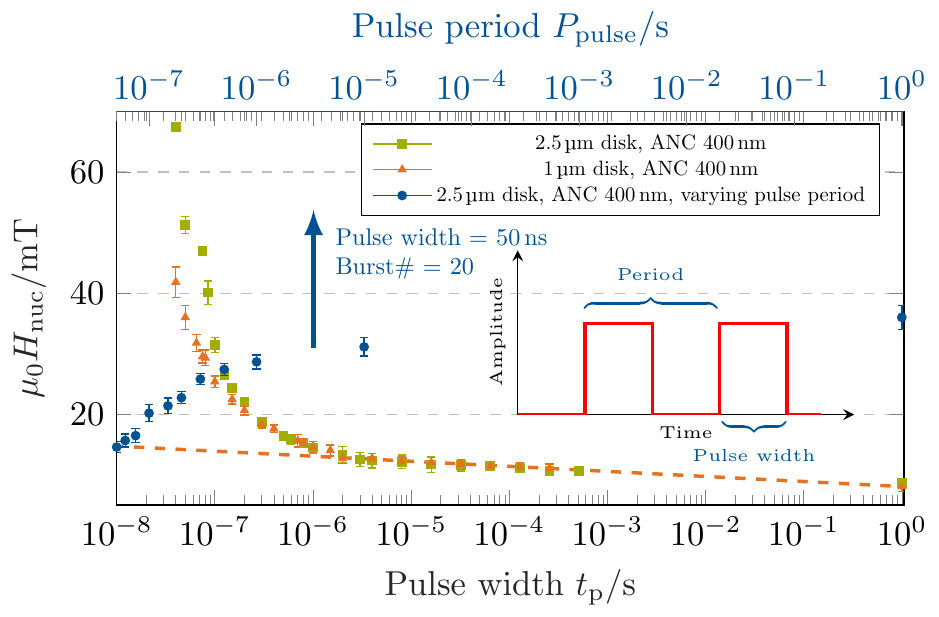}
\caption{Combined plot, showing on the lower \textit{x}-axis the nucleation fields over different pulse widths, \SI{1}{\micro\meter} (orange) and \SI{2.5}{\micro\meter} (green) nano-disks. The plot associated with the upper \textit{x}-axis (blue) displays a sweep of the pulse period $P_{\mathrm{pulse}}$ and its effects on the measured nucleation fields. The cumulative pulse-width is thereby kept constant at \SI{1}{\micro\second} ($ \mathrm{Burst} \# = \frac{\SI{1}{\micro\second}}{\SI{50}{\nano\second}} ).$}
\label{fig:Depinn_proof}  
\end{figure}

\begin{figure}[t]
\centering
\includegraphics[width=0.5\textwidth]{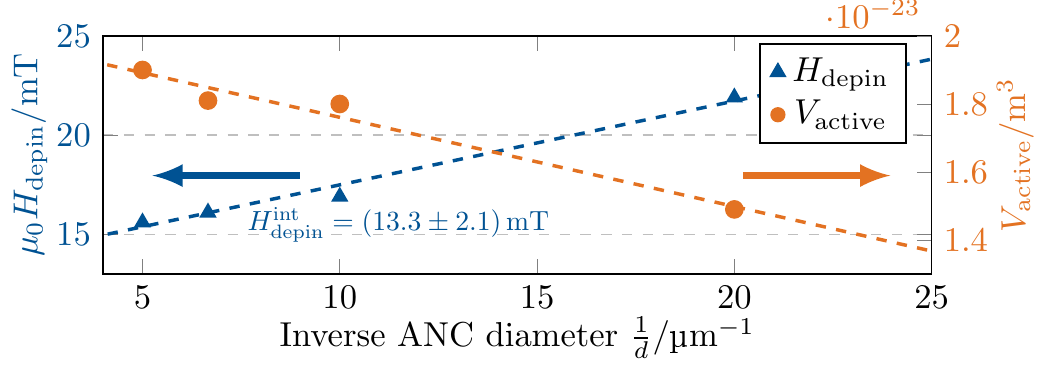}
\caption{A Plot of the depinning fields $H_{\mathrm{depin}}$ at \SI{0}{\kelvin} (on the left) combined with the activation volume $V_{\mathrm{a}}$ (right) for the depinning from a circular ANC, depending on the ANC curvature $\frac{1}{d_{\mathrm{ANC}}}$. Both exhibit a linear $\frac{1}{d}$ dependence, however, with complementary slopes. The dashed lines depict the best linear fits. }
\label{fig:Depinn_radius}  
\end{figure}

Controlling the position of DW nucleation with high spatial accuracy is an essential requirement for prospective DW applications. By targeted irradiation, the anisotropy can, in principle, be lowered locally, creating so-called artificial nucleation centers (ANC) \cite{breitkreutz2012controlled}. However, the known occurrence of significant anisotropy lowering (with unknown distribution) towards the edges severely impedes efforts to create the nucleation volume with the lowest PMA reliably. Nucleation by DW depinning from a fixed anisotropy gradient (e.g., an area with strongly reduced or easy-plane (negative) anisotropy), however, offers an interesting alternative. Here, the anisotropy can be lowered by much larger extents, provided that the depinning fields fall below the intrinsic nucleation fields (via coherent rotation) \cite{franken2011domain}. 
\begin{figure}[b]
\centering
\includegraphics[width=0.5\textwidth]{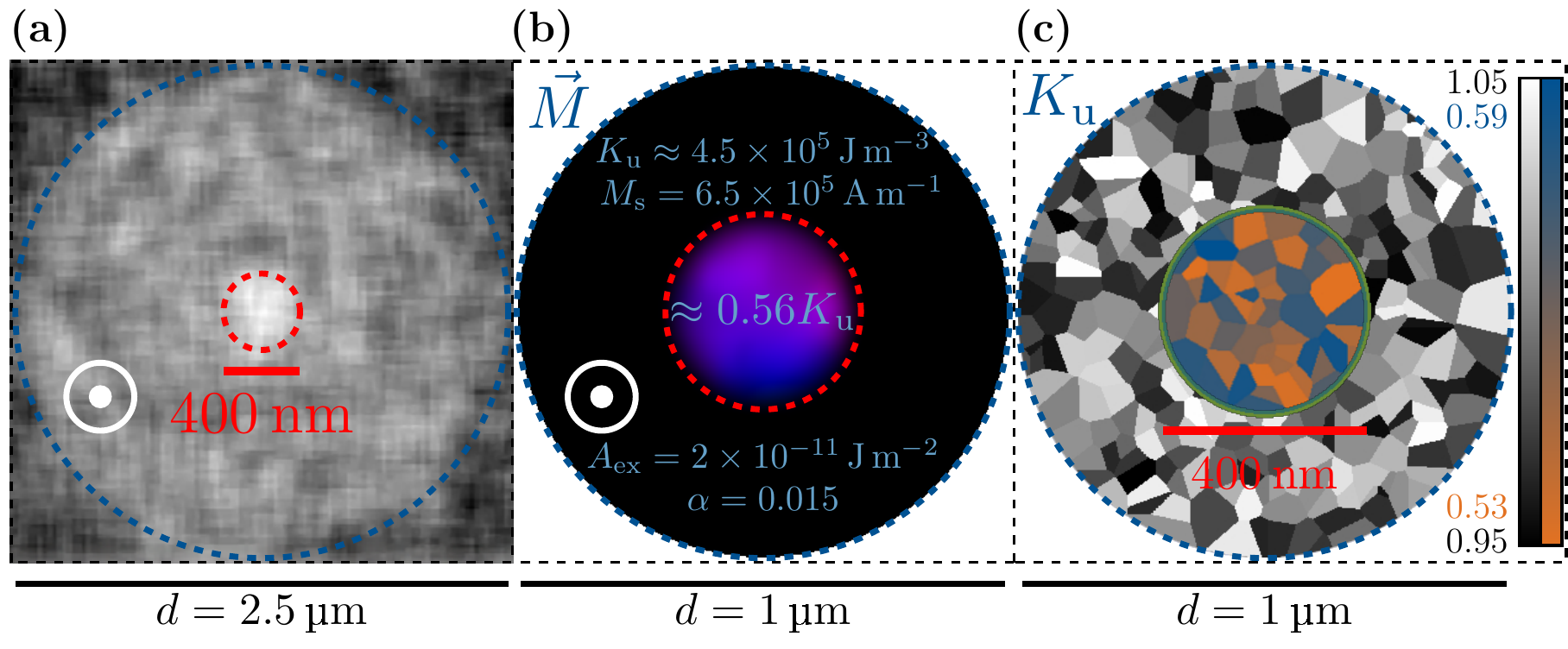}
\caption{Image (a) depicts a differential WMOKE image of a Ta/CoFeB/MgO nano-disk with a diameter of \SI{2.5}{\micro\meter}. The image displays the remanent magnetization at \SI{0}{\milli\tesla} after saturation with \SI{10}{\milli\tesla}. The ANC with a diameter of $\approx \SI{400}{\nano\meter}$ is visible at the center, with its magnetization seemingly pointing opposite to the remaining magnet. Image (b) displays the domain configuration of the respective MuMax3 model with a diameter of \SI{1}{\micro\meter} in remanence. The material parameters are depicted, with special notice, given to $K_{\mathrm{u}}$ inside the ANC area. Image (c) depicts the grain structure of one of the simulated samples, with the colors representing the respective anisotropies. ANC and magnet are separated by a  $\SI{30}{\nano \meter}$ broad transition region (illustrated in green) with a linear anisotropy gradient. The simulated dots' grain and mesh sizes are set to $\approx \SI{15}{\nano \meter}$ and $2.5 \times \SI{2.5}{\nano\meter}$, respectively. }
\label{fig:ANC_Point}  
\end{figure}
Furthermore, the depinning process is governed by different time dynamics, leading to potentially lower switching fields upon approaching timescales close to $\tau_{\mathrm{0}}$, which are, of course, most interesting for applications. For this purpose, ANCs with an anisotropy close to zero are placed in the nanomagnets' center ($d = \SI{1}{\micro\meter}$), employing a double-irradiation approach. First, a homogeneous background irradiation with a dose of $\SI{4,25E13}{}\, \mathrm{ions/cm^2}$ is used to increase $K_{\mathrm{eff}}$ beyond its peak (at $\approx \SI{3.5E13}{}\, \mathrm{ions/cm^2}$). The effective anisotropy is subsequently reduced by a second, target irradiation in the center, with an additional $\SI{3.8E13}{}\, \mathrm{ions/cm^2}$ leading to a cumulative total dose of $\approx \SI{8E13}{}\, \mathrm{ions/cm^2}$ for the ANC. For this dose, Fig.~\ref{fig:HvsKeff} (b) shows a coercivity of $\approx \SI{0}{\milli\tesla}$ with the magnetization effectively following the external field. The ANC position and magnetization direction in remanence is observed by differential WMOKE imaging of larger \SI{2.5}{\micro\meter} wide magnets as displayed in Fig.~\ref{fig:ANC_Point} (a). The bright spot ($d \approx \SI{400}{\nano\meter}$) at the center of the circular magnet, which matches the irradiated ANC area's size, indicates a change in the magnetization direction. However, it is not clear whether the magnetization of the ANC points in-plane or whether it is being aligned anti-parallel by the demagnetizing fields of the host magnet. Complementary to the experiments, a simulation model was developed to better analyze and understand the magnetization reversal in this geometry. The model parameters are chosen to best approximate the characterized magnets. A detailed representation is depicted in Fig.~\ref{fig:ANC_Point} (b,c) (see SI for
more additional information). The depinning from the ANC can be verified by analyzing the time dependence of the switching fields. This is done for a series of magnets with centered circular ANCs (diameters ranging from $d =\SI{100}{\nano\meter}$ to $d =\SI{400}{\nano\meter}$). The cumulative ion dose of all ANCs is $\SI{8E13}{}\, \mathrm{ions/cm^2}$ (keeping in mind the background dose of $\SI{4,25E13}{}\, \mathrm{ions/cm^2}$). 
\begin{figure}[b]
\centering
\includegraphics[width=0.5\textwidth]{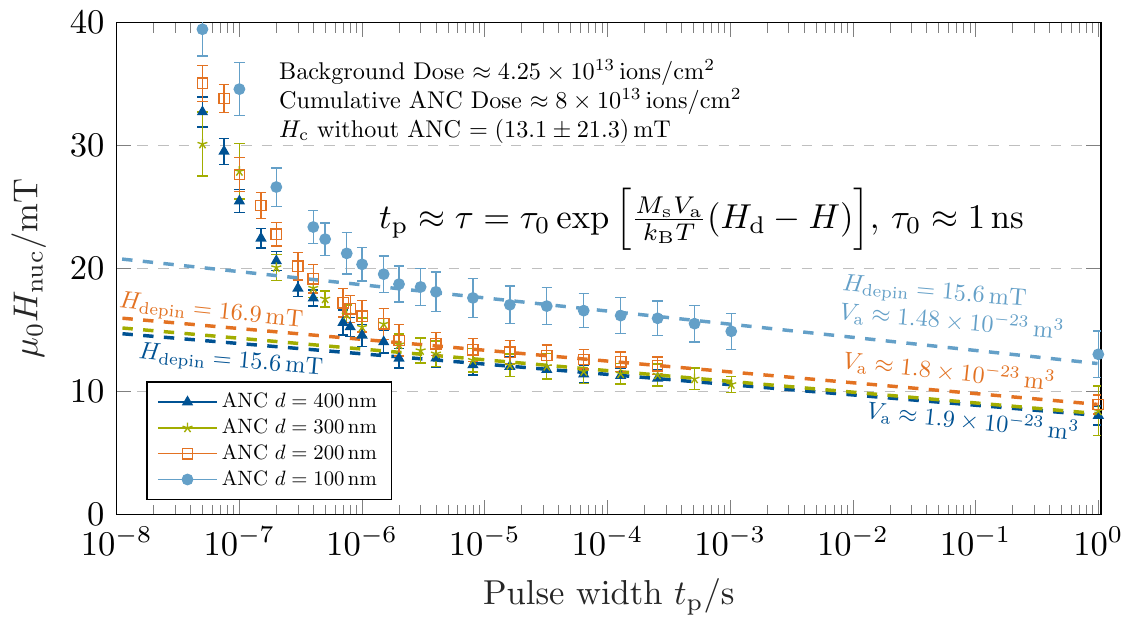}
\caption{Measured nucleation fields of double-irradiated nano-disks ($d = \SI{1}{\micro\meter}$) as a function of the applied pulse width $t_{\mathrm{p}}$. The disks feature circular, different sized ANCs at their centers with the respective diameter given in the legend.}
\label{fig:Depinn}  
\end{figure}
Fig.~\ref{fig:Depinn} depicts the measured nucleation fields with their corresponding fits according to equation~\eqref{Eq:Depinn}. The measured nucleation fields appear to agree well with the depinning model down to low \SI{}{\micro\second} timescales. From this point onward, $H_{\mathrm{nuc}}$ seemingly increases drastically, reaching levels close to those of the irradiated magnets in Fig.~\ref{fig:sharrock}. However, a doubling of the nucleation fields within one order of magnitude (time) is hardly explainable by any reasonable depinning or rotation model. To explain the observed increase in $H_{\mathrm{nuc}}$, it is necessary to consider the measurement procedure discussed in section \ref{sec:Imaging}. After the initial (\SI{}{\nano\second}-long) nucleation pulse, a secondary (\SI{}{\milli\second}-long) low field pulse is used to propagate the DW and ensure a complete magnetization reversal. However, the time between these two pulses allows the magnetization to relax back into the nearest local energy minimum. For a significant portion of the reversal process, this means to flip back into the initial state. We attempt to explain this phenomenon by a simplified but vivid model and underline it via micro-magnetic simulations and related measurements. After the initial depinning from the ANC, the domain expansion can, in first approximation, be modeled as the expansion of a circular bubble from the point of depinning (engulfing half of the ANC area to reduce its DW length). During this process, the system gains exchange and anisotropy energy as the DW length grows with the circumference ($ \propto 2\pi r_{\mathrm{domain}}$) until reaching the magnet's edge, where it splits into two DWs with lengths $\propto r_{\mathrm{magent}}$. The reducing demagnetizing fields do not compensate for this energy gain, as the magnet features a single-domain ground state. Without an external field, the bubble provided it has not reached the edge tends to collapse (it snaps back to the starting point) as the DW tries to lose energy by reducing its length. This effective force on the DW is also described as a Laplace-like pressure, reported in circular domain-structures, with a $\frac{1}{r}$ dependence \cite{moon2011long, zhang2018domain, zhang2018direct}. Figure~\ref{fig:Laplace} illustrates the evolution of the total energy (without \textit{Zeeman} terms) and respective snap-shots of the domain structure throughout the reversal process. Data and images are derived from MuMax3 micromagnetic simulations of a \SI{1}{\micro\meter} nanomagnet with a centered ANC according to Fig.~\ref{fig:ANC_Point} (b,c) \cite{MuMax3}. The simulation parameters (listed in the plot) are thereby chosen to resemble the characterized magnets best. As described in the model above, the total energy initially increases significantly as the bubble domain expands towards the edge, where it reaches a tipping point before falling off, as the DW splits, reducing its length. After overcoming this energy barrier, the domain configuration can be described as quasi-stable until the propagation pulse completes the reversal process. In other words, Fig.~\ref{fig:Depinn} displays the fields necessary to form a quasi-stable domain rather than to depin a DW.
In addition to dynamic simulations, it is possible to test the model implicitly by measuring certain dependencies. Assuming correctness of the model, larger magnets would require stronger fields to propagate the DW to the edge within the pulse duration. Figure~\ref{fig:Depinn_proof} compares the nucleation fields of two different magnet sizes with diameters of \SI{1}{\micro\meter}  and \SI{2.5}{\micro\meter}. For pulse-widths $t_\mathrm{p} < \SI{200}{\nano\second}$, the measured nucleation fields start to diverge, with the larger magnets requiring  significantly higher field strengths for the DWs to form the necessary quasi-stable multi-domain state. However, it has to be noted that data for the \SI{2.5}{\micro\meter} magnets is only available for 3 samples, compared to the 40 for the \SI{1}{\micro\meter} magnets. Besides the dimensional scaling, it is also worth considering the timescales of a possible bubble collapse. Although it is not directly possible to observe this process via WMOKE imaging, information about the timescales at which these collapses occurs can nevertheless be inferred using consecutive on-chip field pulses with varying pulse periods (dead-times between pulses). Using a fixed pulse width of $\SI{50}{\nano\second}$, but sweeping the time between the pulses and measuring the effects on the nucleation fields, it is possible to derive upper and lower bounds for the collapse times. In case the domain collapses within the time between pulses, the switching fields should be independent of the number of pulses (at least in first approximation, not considering the higher attempt count per measurement). Starting at $\approx \SI{1}{\micro\second}$ a clear reduction in the measured nucleation fields is observed. At pulse periods of $\SI{200}{\nano\second}$ ($\SI{150}{\nano\second}$ dead time), however, the measured fields are still twice as high as expected for the cumulative pulse duration of $\SI{1}{\micro\second}$. Only for dead-times  $<\SI{50}{\nano\second}$, comparable nucleation fields are observed.
 All these observations and simulations let us assume that the depinning fields scale according to equation~\eqref{Eq:Depinn} even below $\SI{}{\micro\second}$ pulse widths.\\
Upon analyzing the ANC size-dependent depinning fields in Fig.~\ref{fig:Depinn} and Fig.~\ref{fig:Depinn_radius} , it becomes evident that the depinning process from the circular sources scales $\propto \frac{1}{d_{\mathrm{ANC}}} $ (the curvature of the circle) and thus similar to DW depinning from a notch \cite{kim2009depinning, kim2009analytic, goertz2016domain}. Figure~\ref{fig:Depinn_radius} depicts both the effective activation volumes ($V_{\mathrm{a}}$) and the depinning fields at \SI{0}{\kelvin} versus $ \frac{1}{d_{\mathrm{ANC}}}$. $V_{\mathrm{a}}$ is calculated from equation~\eqref{Eq:Depinn} assuming $M_{\mathrm{s}}\approx \SI{6E5}{\ampere\per\meter}$.
The intrinsic depinning field $H_{\mathrm{depin}}^{\mathrm{int}}$ of the anisotropy gradient can be derived from the zero-intercept of the linear fit to be $H_{\mathrm{depin}}^{\mathrm{int}}= \SI[separate-uncertainty = true]{13.3(21)}{\milli\tesla}$ \cite{kim2009depinning}. 
Analyzing the evolution of the activation volume is more complicated. First of all, it is necessary to point out that the calculated absolute values strongly depend on the value of $M_{\mathrm{s}}$, which is not precisely known. The sizes for $V_{\mathrm{a}}$, although showing a linear $\frac{1}{d_{\mathrm{ANC}}}$ dependence, shrink only marginally compared to the physical dimensions of the respective ANCs. To better illustrate this, we translate the activation volume into an effective ANC diameter $d_{\mathrm{ANC}}^{\mathrm{eff}}$, assuming a cylindrical shaped volume ($d_{\mathrm{ANC}}^{\mathrm{eff}} = 2\sqrt{V_{\mathrm{a}}/(\pi t_{\mathrm{film}})}$). This yields effective diameters from $\approx$ \SIrange{140}{160}{\nano\meter}, indicating that, especially for the larger ANCs, only a small portion takes part in the depinning process. This complies with the depinning models, predicting depinning at the grain with the lowest anisotropy gradient.

\section{Conclusion}
Ta/CoFeB/MgO films and nanostructures were irradiated with Ga\textsuperscript{+} ions to globally and locally modify the magnetic energy landscape, aiming to effectively control the position of DW nucleation.
It has been shown that $ K_{\mathrm{eff}}$ initially increases up to doses of $\SI{3.5E13}{}\, \mathrm{ions/cm^2}$ followed by a steep decline, crossing the easy-plane threshold at $\approx \SI{8E13}{}\, \mathrm{ions/cm^2}$. The time-dependent nucleation field analysis of irradiated magnets revealed shrinking nucleation volumes, despite increases in the anisotropy field and $ K_{\mathrm{eff}}$. Control over nucleation points and fields is achieved, employing a second focused irradiation, creating artificial regions with easy-plane magnetization, from which a DW can depin. The fields needed to depin a DW from this anisotropy gradient scale $\propto \frac{1}{d_{\mathrm{ANC}}} $.

\begin{acknowledgments}
The authors would like to thank Michael Wack for the support in VSM measurements. Furthermore, we would like to thank the IGSSE for its financial support. We gratefully acknowledge the support of the NVIDIA Corporation with the donation of a Titan XP GPU, which was used for this research. Finally, we would like to acknowledge the support of the Central Electronics and Information Technology Laboratory – ZEIT\textsuperscript{lab}.\\
\end{acknowledgments}



\bibliography{bibliography}

\end{document}